\documentstyle[11pt,newpasp,epsf,twoside]{article}
\def\edcomment#1{\iffalse\marginpar{\raggedright\sl#1\/}\else\relax\fi}
\reversemarginpar

\begin{document}
\title{New $\gamma$ Doradus Stars from the Hipparcos Mission and Geneva
Photometry}
\author{Laurent Eyer, Conny Aerts\altaffilmark{1}}
\affil{Instituut voor Sterrenkunde,
             Katholieke Universiteit Leuven,
             Celestijnenlaan 200 B, B-3001 Leuven,
             Belgi\"e}
\altaffiltext{1}{Postdoctoral Fellow, Fund for Scientific Research, Flanders;
e-mail: conny@ster.kuleuven.ac.be}

\begin{abstract}
A search for new $\gamma$ Dor stars was undertaken using
the Hipparcos periodic variable star catalogue and the Geneva
photometric database, leading to a list of 40 new candidates.
We started a monitoring of the candidates which
suited the observational window with the CORALIE spectrograph
at the Swiss Euler Telescope for establishing a robust list
of new $\gamma$ Dor stars and studying line profile variations.
We here present our long-term program.
\end{abstract}

\section{Introduction}
The $\gamma$ Dor stars have amplitude variations up to 0.1 mag
in Johnson V and periods ranging from 0.4 to 3 days (Kaye, these proceedings).
We searched in two databases for finding new members of this class of
variable stars.
The first one is the Hipparcos main mission photometric database.
It contains a mean of 110 measurements for 118\,204 stars brighter
than 12.4 and is magnitude complete up to 7.3-7.9 depending on the
galactic latitude $b$.
As the sampling is ruled by the scanning law of the satellite,
it is not affected by the aliasing around
1/day, which might be a problem for detecting $\gamma$ Dor stars.

The second scanned database is the Geneva photometric catalogue
(Burki \& Kienzle, these proceedings), it counts 48\,000 stars and
345\,000 measurements in a seven colour system.
The content of the Geneva catalogue is the reunion of more than
200 scientific programmes, including namely the Bright Star
Catalogue south of $\delta < +20$.

\section{Hipparcos main mission}
Thousands new variable stars were discovered by the Hipparcos
satellite. During the analysis of the Hipparcos photometry,
stars from the Periodic Catalogue having accurate parallaxes
and colours were plotted in the HR diagram (Eyer 1998).
A clump of stars just at the cool lower edge of the $\delta$
Scuti instability strip was present and gave rise to a
list of 15 candidates (excluding redundant cases from
the other studies). A clump is also present when plotting
the variable stars of the Hipparcos Unsolved Catalogue. 

\section{Hipparcos main mission and Geneva photometry}
A systematic search for finding new $\gamma$ Dor stars
was undertaken in the Hipparcos periodic variable star
catalogue using also Geneva photometry and performing
a multivariate discriminant analysis. This study led to a list
of 14 new $\gamma$ Dor stars (Aerts et al. 1998).
This method is stricter since information on amplitude,
period, physical parameters and multiperiodic behaviour were taken into
account.

\section{The search in Geneva photometry}
Finally the Geneva photometric database was scanned to find
F dwarf stars with high standard deviation. Eleven candidates
were then measured with the 70-cm Swiss telescope, resulting in
about 1000 photometric measurements, which are under study
(Eyer \& Aerts, in preparation). It turns out that about half
suspected stars might be constant stars.

\section{Spectroscopic measurements}
In order to confirm the pulsational character of these stars,
new spectra have been taken with the CORALIE spectrograph on
the 1.2-m Swiss telescope at ESO-La Silla
Observatory. The photometry and spectroscopy are necessary
steps since we want to establish a robust list of new $\gamma$
Dor stars.
Up to now 22 stars have been measured, the strategy consist of
taking at least five spectra of each candidate.
Among the stars, some are binaries, some are fast rotators
and some show clear line profile variations (cf. Fig.~1).
Some stars are too faint for the telescope size, thus
correlation techniques are used to lower the noise level.
The following step is to accumulate photometry and spectra
for promising candidates in order to perform mode identification.
\begin{figure}[thb]
  \plotfiddle{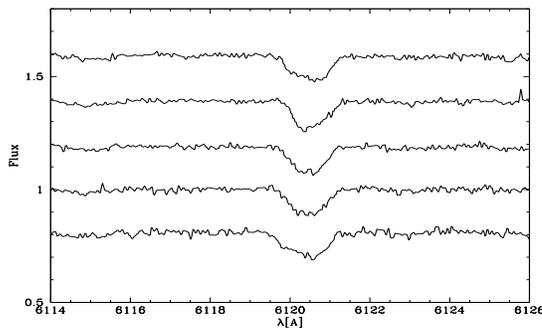}{3.6cm}{-90}{28}{23}{-120}{120}
  \caption{Line profile variations of the star HD~14940.} 
\end{figure}

\end{document}